\documentstyle[prl,aps,epsfig,multicol]{revtex}
\begin{document}
\title{Coexistence of antiferromagnetism and dimerization in a 
disordered spin-Peierls model: exact results}
\author{Michele Fabrizio$^{(a,b)}$, and R\'egis M\'elin$^{(a)}$} 
\address{$^{(a)}$ International School for Advanced Studies SISSA-ISAS, Via
Beirut 2-4, 34013 Trieste, Italy}
\address{$^{(b)}$ Istituto Nazionale di Fisica della Materia INFM}
\date{12 December 1996}
\maketitle
\begin{abstract}
A model of disordered spin-Peierls system is considered, where 
domain walls are randomly distributed as a telegraph noise. 
For this realization of the disorder
in an XX spin chain, we calculate exactly the density of states as well as
several thermodynamic quantities. The resulting physical
behavior should be qualitatively unchanged even for an XXZ chain, up 
to the isotropic XXX point. For weak disorder,
besides a high energy regime where the behavior of a pure spin-Peierls system
is recovered, there is a cross-over to a low energy regime with singular 
thermodynamic properties and enhanced antiferromagnetic fluctuations. 
These regimes are analyzed with the help of exact results, and the relevant 
energy scales determined. We discuss the possible relevance of such a
disorder realization to the doped inorganic spin-Peierls compound CuGeO$_3$.

\end{abstract}
\begin{multicols}{2}
One dimensional quantum spin systems in the presence of randomness show 
unusual and intriguing properties (see e.g.  
Ref. \cite{Fisher}, and references therein). 
For instance, it has been shown\cite{Ma,Hirsh,Fisher} that the ground state of 
the Heisenberg antiferromagnet with
random exchange constants 
can be interpreted as a random singlet state,
where pairs of spins are coupled into singlets with an energy gap to the
triplet configuration which is weaker for widely separated pairs.
The uniform and staggered magnetic susceptibilities, $\chi$ and $\chi_s$,
have a (Griffith's like) singular behavior at low temperature, 
$\chi \sim \chi_s \sim 1/(T\ln^2T)$. 
Interestingly, in spite of the singlet nature of the ground state,
the spin-spin correlation functions are still long-ranged. In fact,
$\langle S^i(r)S^i(0) \rangle \simeq 1/r^2$,
for large $r$ ($i=x,y,z$)\cite{Fisher}. These properties are not modified
by spin anisotropy if, on average, $J_z\leq J_x=J_y$. Notice that
spin anisotropy does not manifest
itself in the spin-spin correlation function with different power law 
behavior of $i=z$ with respect to $i=x,y$, contrary to the case in the 
absence of disorder. The behavior of the random XXZ chain
is however unstable towards a finite average dimerization, i.e. 
a finite average difference between 
the exchange constants of the even bonds and of the
odd bonds. This case 
was recently analyzed by Hyman {\it et al.} \cite{Bhatt} by means of a 
real space renormalization group approach.
For a finite average dimerization $\phi$, 
they find that the spin-spin correlation functions decay 
exponentially with a correlation length $\xi\sim |\phi|^{-2}$, but
the Griffith singularities remain, even if weaker.
In particular, singularities of the uniform susceptibility 
$\chi \sim T^{\alpha -1}$, and the specific heat $C_v \sim T^\alpha$,
where $\alpha\propto |\phi|$, are found to persist\cite{Bhatt}.

The study of the role of disorder in a spin-Peierls system may
be useful to understand the behavior upon doping of the inorganic
spin-Peierls compound CuGeO$_3$. The pure compound is known to undergo
a structural transition at 14K\cite{sp-transition}, below which the CuO$_2$
chains dimerize and a spin-gap opens. However, upon substitution of 
few percent of Cu with magnetic (Ni \cite{Ni}) or non 
magnetic (Zn \cite{Zn,exp}) 
impurities (as well as replacing Ge with Si \cite{Si}), besides the 
structural transition, which still
occurs close to 14K, an antiferromagnetically ordered phase appears 
below a lower temperature $T_N\sim 4K$. Moreover,
the estimated magnetic moment with $4\%$ of Zn is as high as 
0.2$\mu_B$\cite{exp}. This behavior is quite puzzling. First of all,
heuristically, one would expect a N\'eel temperature exponentially small
in the ratio of the average distance between the impurities to 
the spin-Peierls correlation length $\lambda_{SP}$. 
At $4\%$ doping, this would 
imply $T_N/T_{SP}\simeq 0.04$, inconsistent with the experiment. In addition,
one would also expect a magnetic moment of the order of the doping
concentration, not almost an order of magnitude larger, as seen experimentally.
        
In this Letter, we study a particular realization of a disordered
spin-Peierls system which does show a large enhancement of antiferromagnetic 
fluctuations, coexisting on a lower energy scale with an underlying
dimerization. Moreover, this model permits an exact 
calculation of physical quantities for a wide range of temperature/energy.

The Hamiltonian of each chain in the absence of impurities is
\begin{equation}
\hat{H} = \sum_i (1 + \phi_0 (-1)^i) \left( S^x_{i}S^x_{i+1} +
S^y_{i}S^y_{ i+1} + \Delta S^z_{ i}S^z_{ i+1} \right),
\label{H}
\end{equation}
where $\phi_0$ is the strength of the dimerization.  
We assume that one impurity releases one spin-1/2 solitonic excitation,
connecting regions of different dimerization parity\cite{notaSi}. The role
of the interchain coupling is to provide a confining potential to the 
soliton, which will be trapped within some distance from the 
impurity\cite{Khomskii}. Moreover,
the weak link connecting the impurity nearest neighbors (which would be
for instance generated by a next-nearest neighbor exchange) is  
approximated to be equal to the weak bonds in (\ref{H}).
Therefore, the effective Hamiltonian, 
defined now on a chain of one site less,
remains the same apart from the presence of a domain wall. 
For a finite number $n_{imp}$ of randomly distributed impurities, 
the effective model will therefore be assumed to consist of a chain 
with $n_{imp}$ sites less, described by the
same Hamiltonian Eq.(\ref{H}), but in the presence of 
randomly distributed domain walls. This amounts to take
a site dependent $\phi(i)$, which takes alternatively two values 
$\pm \phi_0$, jumping from one to the other at the (random) position
of the antiphase walls. We will show that it is possible to
calculate many physical properties of the soliton band which is
created by disorder inside the spin-Peierls gap, without the
precise knowledge of the soliton wave functions.          
In  Eq.(\ref{H}), $\Delta=0$ corresponds to the XX chain, while $\Delta=1$
is the isotropic XXX model. On the basis of the analyses of 
Refs.\cite{Hirsh,Fisher,Bhatt}, we expect that the behavior at $0<\Delta\leq 1$
should be similar to that at $\Delta=0$, therefore we will only study
the latter case, which is much simpler. We believe that this approximation
gives qualitatively good results for all the range $0\leq\Delta\leq 1$,
especially in view of our particular choice of the disorder. 
By means of a Jordan-Wigner transformation, the model can be mapped
onto a model of disordered spinless fermions.   
By linearizing the spectrum around the
Fermi energy, introducing the right and left moving 
components of the fermion field, and then taking the continuum limit,
the diagonalization of the Hamiltonian amounts to solve the following
coupled differential equations:
\begin{eqnarray*}
-i\frac{\partial}{\partial x}\chi_{R\epsilon}(x) + 
\phi(x)\chi_{L\epsilon}(x) + ih_s \chi_{L\epsilon}(x) &=& 
\epsilon \chi_{R\epsilon}(x), \\
i\frac{\partial}{\partial x}\chi_{L\epsilon}(x) + 
\phi(x)\chi_{R\epsilon}(x) -ih_s \chi_{R\epsilon}(x)&=& 
\epsilon \chi_{L\epsilon}(x), 
\end{eqnarray*}
where $\chi_{R(L)\epsilon}(x)$ is the eigenfunction of
energy $\epsilon$ on the right(left) moving field, and we have also 
considered for later convenience
a uniform staggered magnetic field $h_s$ in the $z$-direction.
The dimerization field $\phi(x)$ corresponds
to that introduced in Eq.(\ref{H}), apart from an appropriate
normalization factor.  
The equations can be decoupled by the following transformation
\begin{eqnarray*}
u_{+\epsilon}(x) &=& \chi_{R\epsilon}(x) + i\chi_{L\epsilon}(x),\\
u_{-\epsilon}(x) &=& i\chi_{R\epsilon}(x) - \chi_{L\epsilon}(x).
\end{eqnarray*}
These two functions are solutions of the Schr{\oe}dinger-like equations
\[
\left( -\frac{\partial^2}{\partial x^2} + \phi^2(x) \pm \phi'(x)\right)
u_{\pm \epsilon}(x) = E u_{\pm\epsilon}(x),  
\]
where $E=\epsilon^2 - h_s^2$ should be greater than zero. 
In the following, we will often use the integrated density
of states as a function of $E$, which we will define as $N(E)$.
In terms of this function, the density of states of the fermionic
model is 
\begin{equation}
\rho(\epsilon) = 2\epsilon 
\frac{\partial N(E)}{\partial E}_{|_{\scriptstyle E=\epsilon^2 - h^2_s}}.
\label{N}
\end{equation}
In the case in which $\phi(x)$ is a white noise,
these equations have been analyzed quite in detail in the context of
disordered one-dimensional Fermi systems \cite{Ovchinnikov,Eggarter}, or 
classical diffusion of a particle in a random medium (for a review 
see e.g. Ref.\cite{Bouchaud}). 
An interesting anomaly of this problem is that, for a 
zero-average white noise, the $E=0$ state is extended 
\cite{Eggarter,Pastur,Tosatti}, and both the localization length
and the density of states diverge as $\epsilon\to 0$.
Quite recently, Comtet, Desbois and
Monthus\cite{Comtet} (CDM) specialized those equations for a particular 
disorder, for which they have been able to calculate
exactly the integrated density of states $N(E)$ and the localization 
length $\lambda(E)$.
Specifically, they assumed a random potential $\phi(x)$ 
which takes alternatively two values
$\phi_0$ and $\phi_1$ at intervals whose lengths $l\geq 0$ are randomly
distributed according to the probability densities 
$f_0(l) = n_0 \exp{(-n_0 l)}$ and $f_1(l) = n_1 \exp{(-n_1 l)}$
(see also Ref.\cite{Pastur}).
This choice of $\phi(x)$ is particularly suited for studying our
problem of randomly distributed domain walls. 
In particular, our case corresponds to $\phi_1=-\phi_0<0$,
and $n_0 = n_1$, i.e. to an average dimerization 
$\phi = (\phi_0 n_1 + \phi_1 n_0)/(n_0+n_1) = 0$. Nevertheless, we will 
also discuss the more general situation $n_0\not= n_1$, in which case
$\phi$ is finite.  
Moreover, we start by taking $h_s=0$.
In the model there are three 
relevant length scales, $\lambda_{SP} = 1/\phi_0$, $l_0=1/n_0$
and $l_1=1/n_1$. $\lambda_{SP}$ is the correlation length of
the system in the absence of disorder, which is the case if 
for instance $l_0/l_1 \to \infty$. In this case, the spectrum of the 
single-particle excitations (which is symmetric around zero
energy) shows a gap $2\phi_0$, and the density of states $\rho(\epsilon)$ has
an inverse square root singularity at $\epsilon=\pm \phi_0$.   
For generic $l_0$ and $l_1$, the density of states can still be
exactly calculated within the phase formalism approach \cite{Comtet}, and
expressed in terms of integrals which have to be numerically evaluated. 
Essentially, the method consists in writing the master equation for
the joint probability distribution of the phase of the wave function and
$\phi(x)$, and solving for the stationary $x$-independent solution.
In particular,
if $l_0$ and $l_1$ are much longer than $\lambda_{SP}$, i.e. if
the gap has the time to develop in a region of constant $\phi(x)$,
the density of states still shows a peak at $\pm \phi_0$, even though 
states are created inside the gap. 
These states accumulate, in a singular manner, as $\epsilon\to 0$
In particular, the density of states
around zero energy goes like $\rho(\epsilon)\sim \epsilon ^{2\mu -1}$,
where $\mu = (n_1-n_0)/(2\phi_0)$ is finite. 
In Fig.1, we draw
$\rho(\epsilon)$ for $\epsilon>0$, $\phi_0=1$, and various $n_0$ and $n_1$. 
For $\mu=0$, $\rho(\epsilon)\sim 1/|\epsilon\ln^3\epsilon|$.
The key feature of our choice for the random potential is that,
even if the average dimerization 
$\phi=0$,
i.e. if $n_0=n_1=n$, the density of states shows a pseudo-gap
if $\phi_0\gg n$ (see Fig.1),  
totally absent for a white noise process\cite{Xu}.

To be more precise, from our numerical results we find, similarly to CDM, that
the integrated density of states $N(E)$ 
for weak disorder (i.e. both $n_0$ and $n_1$ much smaller than
$\phi_0$) saturates below the
pseudogap $\phi_0$ to a value $N_*\sim n_0n_1/(n_0+n_1)$, which is
of the order of half the average number per unit length of steps of the 
random potential. The saturation occurs at an energy scale $E_*$ which can be
identified as the typical effective bandwidth of those midgap excitations.
This result physically implies that, for weak disorder, the number of states
generated inside the gap is of the order of the average number of
domain walls. From the analytical expression
of $N(E)$, we obtain that $\ln (E_*/\phi_0^2) \sim - 2\phi_0/(n_0+n_1)$, 
i.e. $E_*$ is exponentially small in the inverse of the disorder strength. 
In addition, it is also possible to calculate the localization length
$\lambda(E)$. In particular, for $\phi\not =0$, 
$\lambda(0)=1/\phi$, which implies that the localized wavefunctions inside the 
gap have a much longer localization length 
than the spin-Peierls correlation length 
$\lambda_{SP}$. More interesting, for $\phi=0$, which is relevant 
for our disorder modelization,
$\lambda(E)\sim |\ln E |$, so that the states close
to $E=0$ are almost delocalized. 

More generally, our 
model at low temperature/energy is equivalent to the models
analyzed in Ref.\cite{Fisher} and in 
Ref.\cite{Bhatt}, for $\phi=0$ and $\phi\not=0$, respectively. The
analogy can be expected by the following arguments. 
For $\epsilon\leq \phi_0$ and $n_0=n_1\ll \phi_0$,
the problem
reduces to a model of weakly coupled spins localized close to each domain wall.
As a first approximation, only the exchange coupling between two successive 
spins can be retained, which is given by   
$J(r)\simeq \phi_0 \exp(-r\phi_0)$, being $r$ the random distance between 
two domain walls distributed according to $n\exp(-rn)$.
Thus the model is indeed equivalent to an Heisenberg chain with
randomly distributed exchange constants. 
The probability distribution of $J$ at energy scales $\leq \phi_0$ 
can be readily found to be
\[
P(J) = \theta(\phi_0 - J)\left(\frac{n}{\phi_0^2}\right)
\left(\frac{\phi_0}{J}\right)^{1-n/\phi_0},
\]  
and it has to be used as the starting point of the renormalization
group flow equations of Ref.\cite{Fisher}. In this way, it is possible to 
recover the same results that we obtain by exploiting the exact solvability
of our model, thus showing not only that the two models are equivalent,
but also that spin-anisotropy does not really matter \cite{noi}.  
For $n_0\not= n_1$, the
same analogy works now with the model of Ref.\cite{Bhatt}.
More rigorously, the above conjectured equivalence can be proven by showing 
that the models have the same low temperature thermodynamic properties.
 
In our model, we can in fact calculate exactly many thermodynamic quantities 
and find not only the low temperature but also the intermediate 
($T\sim \phi_0$) temperature behavior. For instance, the uniform magnetic 
susceptibility is given by 
\[
\chi(T) = \beta \int_0^\infty dE \frac{\partial N}{\partial E}
\frac{1}{2\cosh^2(\beta\sqrt{E}/2)},
\]
and is plotted in Fig.2 for the same values of $\phi_0$, $n_0$ and $n_1$
as in Fig.1. 

From the asymptotic behavior of $N(E)$ for small $E$,
we find that, at low $T$, $\chi(T) \sim T^{2\mu-1}$,
for $\mu\not=0$, and $\sim 1/(T\ln^2T)$ for $\mu=0$. The latter is
exactly the result for the random XXZ Heisenberg model.
Our model thus belongs, at low energy and for $\mu=0$, to the same
universality class. For all $\mu$'s smaller than $1/2$, the
magnetic susceptibility still diverges at low temperature.
Analogously, the specific heat vanishes as $C_v\sim T^{2\mu}$
 ($C_v\sim 1/|\ln^3T|$, for $\mu=0$),
which is compatible with the result of Ref.\cite{Bhatt} with 
$2\mu = \alpha$, thus showing the equivalence with our model at
$\mu\not=0$. In addition, we obtain the full behavior of
$\chi$ at intermediate temperatures, as shown in Fig.2.
We see that, at $T\sim \phi_0$, the susceptibility decreases 
as if a spin-Peierls gap were present, even though it finally diverges
at low $T$. Moreover, for $E_*<T<\phi_0$, we predict a Curie like
behavior, with a Curie constant $\propto N_*$.

The behavior of the staggered part of the spin-spin 
correlation function $\chi_s(x)$ can be deduced by the analogies
with the models analyzed in Refs. \cite{Fisher,Bhatt}.
In particular, for $\phi\not=0$, 
$\chi_s(x)$ decays exponentially
with a correlation length $\propto (1/\mu)^2$\cite{Bhatt}. On the contrary, 
for the case relevant to our model, which corresponds to
$\phi=0$,  $\chi_s(x)$ decays as a power law $\sim 1/x^2$\cite{Fisher}.
At finite temperature and $\mu=0$,  
$\ln \chi_s(x,T) \sim -x \sigma/\ln^2(T/\phi_0)$ \cite{Fisher}, 
where $\sigma = 2\phi_0^2/(n_0+n_1)$. 
This expression suggests a new energy scale $E_c$, which
can be identified as the coherence energy for the antiferromagnetic
fluctuations. In fact, when $T\geq E_c$, the
correlation function should behave like $\exp{(-2x\phi_0)}$, which leads
to $\ln (E_c/\phi_0^2) \sim -\sqrt{\sigma/\phi_0}$, 
that is to a coherence
energy exponentially small in the inverse square root of the disorder strength,
but still much bigger than $E_*$. The appearance of an energy scale
governing the spin-spin correlation function, which differs from that
entering the average density of states, is not unexpected in the 
presence of disorder, which introduces basic differences between average
and typical behaviors. 
On the other hand, for $\mu\not=0$,
below another energy scale $E_\mu$, we should recover the result of
Ref.\cite{Bhatt}, which sets $\ln (E_\mu/\phi_0^2) \sim - 1/\mu$. 

We also exactly calculate
the staggered susceptibility $\chi_s(T)$. 
By means of Eq.(\ref{N}) we find that 
\begin{equation}
\chi_s(T) = \int_0^\infty dE \frac{\partial N}{\partial E}
\tanh\left(\frac{\beta \sqrt{E}}{2}\right) \frac{1}{\sqrt{E}}.
\label{chis}
\end{equation}
For $\mu<1/2$, this susceptibility diverges at low $T$ like the
uniform susceptibility. However, while the integral over $E$ in
the uniform susceptibility is cut-off by $T^2$, the contribution to 
the singular behavior of the staggered susceptibility comes from all
$E$ up to approximately $E_*$. Moreover, all higher energies also contribute
to the staggered susceptibility with a finite term as $T\to 0$. 
Therefore, while the singular behavior
deriving from all $\epsilon = \sqrt{E} <T$  
can be ascribed to local excitations, that deriving from $\epsilon >T$ 
is solely due to longer range antiferromagnetic fluctuations\cite{Fisher}.
The rapid enhancement of antiferromagnetic fluctuations that we find
is extremely suggestive in the light
of that recently observed in CuGeO$_3$, as previously discussed.
In fact, our model for a disordered spin-Peierls system clearly shows a
coexistence of dimerization with long range antiferromagnetic fluctuations,
the latter existing on energy scales lower than the pure spin-Peierls gap.
These fluctuations may induce a magnetic ordering
via the interchain coupling, below some N\'eel temperature $T_N$.
The magnetic susceptibility would then still show the drop
at the Peierls transition, but the low temperature divergence would
finally be cut-off by $T_N$, below which $\chi(T)$ would exponentially
vanish, compatibly with the experimental evidences (see e.g. Ref.\cite{exp}).
Moreover, on the basis of the previous discussion, we expect that
the N\'eel temperature is related to the energy scale governing the 
spin-spin correlation function, that is to 
$E_c$, which is exponentially small in $\sqrt{l/\lambda_{SP}}$,
and therefore larger than the typical bandwidth of the
low energy excitations $E_*$, which is 
exponentially small in $l/\lambda_{SP}$. This difference might be the
explanation of the relatively large N\'eel temperatures found in
the doped CuGeO$_3$.  

It is a pleasure to acknowledge useful discussions with A.O. Gogolin,
A.A. Nersesyan and Yu Lu. A particular thanks to E. Tosatti, who
has been the source of inspiration of this work. This work has been partly
supported by EEC under Contract No. ERB CHR XCT 940438, and by the 
INFM, project HTSC.

\end{multicols}
\newpage
\begin{figure}
\epsfig{file=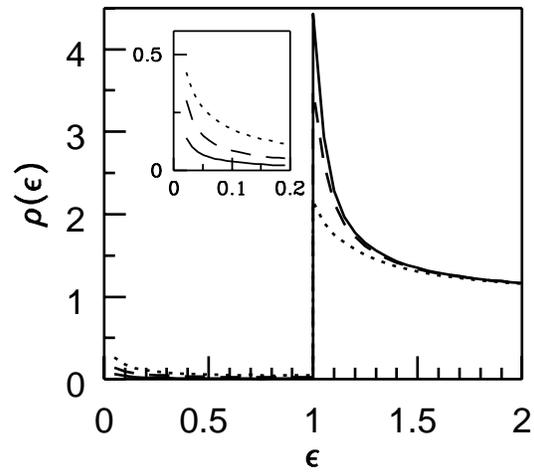}
\caption{ Density of states
for $\phi_0=1$  and  
$n_0=n_1=0.3$ (dotted line),  $n_0=n_1=0.1$ (full line), 
$n_0=0.1$, $n_1=0.3$ (dashed line).
Also shown in the insert is the low energy behavior. }
\end{figure}

\begin{figure}
\epsfig{file=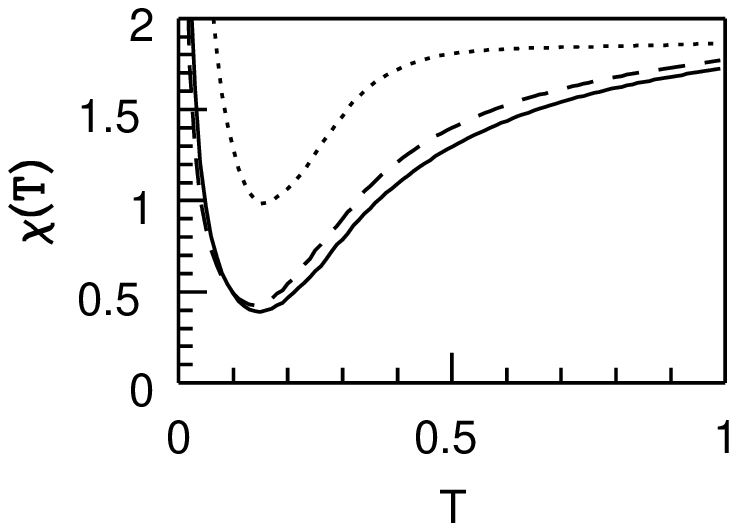}
\caption{ Uniform magnetic susceptibility at zero 
staggered magnetic field,
for the same cases as in Fig.1 }
\end{figure}



\begin{references}

\bibitem{Fisher}D. Fisher, Phys. Rev. B {\bf 50}, 3799 (1994);
{\it ibid.} {\bf 51}, 6411 (1995).
\bibitem{Ma}C. Dasgupta and S.K. Ma, Phys. Rev. B {\bf 22}, 1305 (1980) 
\bibitem{Hirsh}J.E. Hirsh, Phys. Rev. B {\bf 22}, 5339 (1980);
{\it ibid.}, 5355 (1980).
\bibitem{Bhatt}R.A. Hyman, K. Yang, R.N. Bhatt, and S.M. Girvin,
Phys. Rev. Lett. {\bf 76}, 839 (1996).

\bibitem{sp-transition} M. Hase {\it et al.}, Phys. Rev. Lett. {\bf 70}, 
3651 (1993); J.P. Pouget {\it et al.}, Phys. Rev. Lett. {\bf 72}, 4037 (1994);
K. Hirota {\it et al.}, Phys. Rev. Lett. {\bf 73}, 736 (1994).

\bibitem{Ni} J.-G. Lussier {\it et al.} J. Phys. Condens. Matter {\bf 7},
L325 (1995).

\bibitem{Zn} M. Hase {\it et al.}, Physica B {\bf 215}, 164 (1995).

\bibitem{exp} M. Hase {\it et al.}, J. Phys. Soc. Jpn. {\bf 65}, 1392 (1996);
Y. Sasago {\it et al.}, unpublished (1996).

\bibitem{Si} J.-P. Renard {\it et al.} Europhys. Lett. {\bf 30}, 475 (1995);
L.P. Regnault {\it et al.} Europhys. Lett. {\bf 32}, 579 (1995).

\bibitem{notaSi} The assumption that each impurity releases one
soliton is in fact more appropriate to describe the effect of 
Zn or Ni doping. However, there are claims that also Si-doping
can be effectively represented by a random distribution of
domain walls (see T. Ng, cond-mat/9610016).

\bibitem{Khomskii} D. Khomskii, W. Geertsma, and M. Mostovoy, cond-mat/9609244.


\bibitem{Ovchinnikov} A.A. Ovchinnikov, and N.S. \'Erikhman, Sov. Phys.
JETP {\bf 46}, 340 (1977).
 
\bibitem{Eggarter} T.P. Eggarter and R. Riendinger, 
Phys. Rev. B {\bf 18}, 569 (1978). 

\bibitem{Bouchaud}J.P. Bouchaud, A. Comtet, A. Georges, and P. Le Doussal,
Ann. Phys. {\bf 201}, 285 (1990).

\bibitem{Pastur} I.M. Lifshits, S. Gredeskul, and L.A. Pastur,
{\it Introduction to the Theory of Disordered Systems}, 
John Wiley and Sons, New York (1987),

\bibitem{Tosatti}E. Tosatti, M. Zannetti, and L. Pietronero,
Z. Phys. B {\bf 73}, 161 (1988).

\bibitem{Comtet}A. Comtet, J. Desbois, and C. Monthus, Ann. Phys. {\bf 239},
312 (1995) [see also C. Monthus, G. Oshanin, A. Comtet, and
S.F. Burlatsky, Phys. Rev. B {\bf 54}, 231 (1996)].

\bibitem{Xu}B.-C. Xu and S.E. Trullinger, Phys. Rev. Lett {\bf 57}, 3113
(1986), analysed a similar model with a white-noise mass 
by means of a supersymmetric functional-integral formalism.
However, their density of states do not coincide with that exactly 
calculated for instance in Ref.\cite{Ovchinnikov}.
We do not understand the origin of the disagreement.

\bibitem{noi} M. Fabrizio and R. M\'elin, in preparation.


\end{references}
\end{document}